\documentclass[fleqn,12pt,twoside]{article}
\usepackage{espcrc1,psfig,epsfig}

\usepackage{graphicx}
\usepackage[figuresright]{rotating}

\def\spose#1{\hbox to 0pt{#1\hss}}
\def\simlt{\mathrel{\spose{\lower 3pt\hbox{$\mathchar"218$}}
     \raise 2.0pt\hbox{$\mathchar"13C$}}}
\def\simgt{\mathrel{\spose{\lower 3pt\hbox{$\mathchar"218$}}
          \raise 2.0pt\hbox{$\mathchar"13E$}}}

\def\GEtool{{\tt GEtool}}
\def\Ctwelve{$^{12}$C}
\def\Cthirteen{$^{13}$C}

\title{Galactic Chemical Evolution Redux: Atomic Numbers
       6$\le$Z$\le$15}

\author{B.~K. Gibson\address[Swin]{Centre for Astrophysics \& Supercomputing, 
        Swinburne University, \\ 
        Mail \#H31, P.O. Box 218, Hawthorn, Victoria, Australia, 3122}
        \thanks{The financial support of the Australian Research
                Council is gratefully acknowledged.  We likewise
		acknowledge our peerless collaborators in this effort,
		including John Lattanzio, Amanda Karakas, Simon Campbell,
		Alessandro Chieffi, Marco Limongi, Maria Lugaro,
		and Agostino Renda.},
        Y. Fenner\addressmark
	\,and
        A. Kiessling\addressmark}
       
\begin{document}

\maketitle

\begin{abstract}
Motivated by the inability of Galactic chemical evolution models to 
reproduce some of the observed solar neighbourhood distribution of elements (and 
isotopes) with atomic numbers 6$\le$Z$\le$15, we have revisited
the relevant stellar and Galactic models as part of an ambitious new program
aimed at resolving these long-standing discrepancies.  Avoiding the use
of (traditional) parametric models for low- and intermediate-mass stellar
evolution, we have generated a new, physically self-consistent, suite of 
stellar models and integrated the nucleosynthetic outputs into 
\GEtool, our semi-analytical galactic chemical evolution software package.
The predicted temporal evolution of several light- and intermediate-mass
elements (and their isotopes) in the solar neighbourhood 
- from carbon to phosphorus - demonstrate the efficacy of the new yields
in reconciling theory and observation.
\end{abstract}

\section{MOTIVATION}

The role of asymptotic giant branch (AGB) stars in contributing to the
chemical enrichment of the interstellar medium (ISM) has long been appreciated
\cite{RV81,vG97}.
Unfortunately, the dearth of self-consistent grids of non-parametric
stellar models for low- and intermediate-mass stars (LIMS) ensured that
that ``appreciation'' has remained more qualitative than quantitative.
Nowhere has this been more problematic than in attempts to understand
the origin and evolution of the light- and intermediate-mass metals --
atomic numbers between 6 and 15 -- where the discrepency between 
supernovae yields and observed stellar abundances is most acute \cite{TWW95}.

We have recently undertaken an ambitious program of coupled stellar
and Galactic chemical evolution modeling, preliminary results for which 
are described here.  Of order 50 stellar models were run
using the Mount Stromlo Stellar Structure code; a nucleosynthesis
post-processing with time-dependent diffusive mixing was then applied,
in order to derive detailed yield information \cite{KL03a,KL03b}.
The parameter space 
covered was extensive, with masses ranging from 1--7 M$_\odot$,
metallicities ranging from zero to super-solar (for both
scaled-solar and $\alpha$-enhanced abundance ratios), as well as
varying treatments of mass-loss and reaction rates.  Elements for 
which an initial mass function (IMF)-weighted abundance with respect to solar
varied by more than 0.1~dex from that derived using the canonical
LIMS yields of \cite{RV81} include atomic numbers 6$\le$Z$\le$15 (carbon
through phosphorus).

These new stellar yields have now been implemented within \GEtool\ 
\cite{FG03,G03},
a semi-analytical galaxy evolution code which treats the formation
of the Milky Way within a dual-phase infall framework -- the first, a rapid 
infall phase leading to the formation of the stellar halo;
the second, a more extended phase associated with the formation of
the disk.  
Such dual-phase formation scenarios have proven success in reproducing
the metallicity distrubtion of nearby stars \cite{C97,C01}.
\GEtool\ includes a sophisticated treatment of chemical
enrichment in the interstellar medium, with the yields of Type~Ia and 
II supernovae (SNe), Wolf-Rayet stars, and AGB stars, all incorporated.
This powerful combination of \GEtool\ and the new yields has been
demonstrated recently by \cite{F03,F04a,F04b,R04}.

\section{RESULTS}

In the following subsections, we highlight just a few specific examples
for which the new LIMS yields have led to interesting new insights into
the distribution of elemental and isotopic abundances for atomic numbers
6$\le$Z$\le$15.

\subsection{Carbon}

Over the past 4.5~Gyr, the ratio of \Ctwelve-to-\Cthirteen\ (by mass) 
in the local ISM has decreased $\sim$20\%.  Attempts at recovering this 
temporal behaviour in \Ctwelve/\Cthirteen\ has led to the suggestion that 
significant pollution of the local ISM by novae ejecta must have
occurred over the lifetime of the Milky Way's disk \cite{RM03}.
This conclusion was driven in part by the failure of standard LIMS
yields \cite{vG97} to predict the observed decrease in the temporal
evolution of \Ctwelve/\Cthirteen\ over the disk's lifetime - 
see Fig~1 of \cite{RM03}.\footnote{We note that
the ``single star'' \Ctwelve/\Cthirteen\
models of \cite{RM03} were scaled upwards by 35\% in order
to ensure an \it a posteriori \rm agreement with the Sun's
\Ctwelve/\Cthirteen.}
Conversely, after incorporating the new grid of LIMS yields
into \GEtool\ our standard solar neighbourhood model predicts an $\sim$15\%
{\it decrease} in \Ctwelve/\Cthirteen\ over the past 4.5~Gyr, in agreement with
the aforementioned empirical data.  Over the lifetime of the disk, the 
same ratio is predicted to have decreased by $\sim$40\%.  In other words, the
new yields appear to
obviate the need for a putative (significant) nova component
to Galactic chemical evolution.\footnote{The present-day Galactic 
\Ctwelve/\Cthirteen\ gradient is an additional constraint on novae pollution
in the ISM.  We note in passing that our Galactic models, with the new yields,
result in a gradient in excellent agreement with the Galactic 
distribution of carbon monoxide 
\cite{WR94}.  The inclusion of novae (significant
\Cthirteen\ factories) does not impact significantly on the predicted 
gradient in \Ctwelve/\Cthirteen, {\it but} the resulting 
zeropoint is approximately
a factor of two below that of the present-day Galactic ISM.}

\subsection{Fluorine}

The nucleosynthesis pathways for fluorine production in AGB stars
involve both the helium and combined hydrogen-helium burning phases.  The
primary uncertainty in the net production rate can be traced to the adopted
reaction rates of $^{14}$C($\alpha$,$\gamma$)$^{18}$O and 
$^{19}$F($\alpha$,$p$)$^{22}$Ne, and from the inclusion of partial mixing
of protons from the envelope in the top layers of the helium intershell
region \cite{R04}.  Models of Galactic chemical evolution 
which include {\it only} $^{19}$F production from neutrino spallation
of $^{20}$Ne in Type~II SNe underproduce the observed [F/O] in
the Milky Way and LMC by a factor of $\sim$2 \cite{TWW95,R04}.
After including the new LIMS yields within \GEtool, supplemented with the 
$^{19}$F yields for Wolf-Rayet stars at super-solar metallicity
\cite{MM00}, we ran our standard Milky Way
chemical evolution model.  The new yields resulted - for the first
time - in the successful
recovery of the trend of [F/O] observed over $\sim$1~dex in metallicity
in the Milky Way and LMC \cite{R04}.

\subsection{Neon}

The neon isotope $^{22}$Ne is produced with relative ease in the
helium-burning shell of AGB stars via the capture of two $\alpha$-particles
onto residual $^{14}$N remaining from the earlier operation of the 
CN-cycle; subsequent thermal dredge-up mixes this $^{22}$Ne into the
envelope.  This $^{22}$Ne nucleosynthetic pathway appears to operate 
independent of metallicity, but only over a limited mass range
($m$$\approx$3.0$\pm$0.5~M$_\odot$) \cite{KL03b}.
The $^{22}$Ne/$^{20}$Ne ratio for this $^{22}$Ne-enriched ejecta 
(from stars with $m$$\approx$3~M$_\odot$ and $Z$$\simlt$$Z_\odot$) is
$\sim$3, comparable to the ratio expected in Wolf-Rayet stellar
ejecta: $^{22}$Ne/$^{20}$Ne$\approx$3--10 for $m$$\simgt$40~M$_\odot$
and $Z$$\simgt$$Z_\odot$ \cite{HL03}.  The new yields, in combination
with our standard Galactic model, result in a factor of $\sim$2
increase in
$^{22}$Ne/$^{20}$Ne (with no discernible impact on $^{21}$Ne/$^{20}$Ne)
over the past $\sim$7~Gyr history of the solar neighbourhood.
In contrast, the
isotopic ratio $^{22}$Ne/$^{20}$Ne in Galactic cosmic rays has been 
measured to be anomalously high - $\sim$5$\times$ that of the solar wind.
This observation has been attributed to the cosmic rays having been
accelerated from superbubbles of metallicity $\sim$3$Z_\odot$ with the
accompanying Wolf-Rayet (and Type~II SNe) ejecta accounting for 
$\sim$20\% of the local ISM (by mass) \cite{HL03}.  It would be useful
to revisit this latter conclusion in light of this previously unappreciated
source of $^{22}$Ne - AGB stars of mass $\sim$3~M$_\odot$.

\subsection{Sodium}

It is a well-known fact that Galactic chemical evolution models which
incorporate sodium yields from Type~II SNe {\it alone} tend to underproduce
[Na/Fe] by a factor of $\sim$2--3 over $\sim$3 dex in metallicity 
\cite{TWW95}.  It has also been recognised for some time that
the Ne--Na chain acting in AGB stars can lead to
the production of $^{23}$Na via proton capture on $^{22}$Ne
\cite{KL03a}.\footnote{Various nucleosynthesis
pathways exist, including the hydrogen- and helium-burning shells, and
hot bottom burning in massive AGB stars.}  Using our new yields, we have
constructed the first Milky Way chemical evolution model which includes
self-consistently the sodium production from both Type~II SNe and AGB
stars.  Regardless of the Type~II SNe yields adopted, the inclusion of 
sodium from AGB stars results in fairly uniform 0.2-0.4~dex increase in
the predicted [Na/Fe] in the ISM for $-$2$\simlt$[Fe/H]$\simlt$$+$0,
in excellent agreement with observations from \cite{Edv93,Gra03}.
\footnote{It should be noted that
a ``numerical'' (coincidental!)
pseudo-degeneracy exists between sodium from AGB stars and
sodium from stars of mass $m$=40-100~M$_\odot$.  For example, as the most
massive model generated by \cite{WW95} is 40~M$_\odot$, if one decides
to adopt an upper mass limit of, say, 100~M$_\odot$ for the IMF, one 
is forced to extrapolate the sodium production from the lower mass models.
Because this sodium production is a steeply increasing function of 
stellar mass, linearly extrapolating to 100~M$_\odot$ can actually lead
to a predicted [Na/Fe] vs [Fe/H] behaviour which mimics (again,
coincidentally)
that encountered when using the new AGB sodium yields.  A careful 
consideration of very massive star sodium yields must be undertaken before
we can make any further quantitative statements.}

\subsection{Magnesium}

While the bulk of magnesium in the Galaxy can be traced to Type~II SNe, 
\cite{KL03a} have shown that 
sub-solar metallicity AGB stars can be important contributors of 
$^{25}$Mg and $^{26}$Mg isotopes to the ISM.  Nucleosynthesis of these
isotopes is believed to occur
via $\alpha$-capture on $^{22}$Ne triggered by helium shell thermal
pulsing.  More massive AGB stars ($m$=4--6~M$_\odot$) may actually burn
magnesium via hot bottom burning at the base of the convective envelope.
Models of Galactic chemical evolution which include only the heavy magnesium
isotopes returned to the ISM via Type~II SNe are significantly discrepant
with observational data \cite{TWW95}.  Incorporating the Mg isotopic contribution  
from AGB stars into \GEtool\ leads to a factor
of 2--3 increase in $^{25,26}$Mg/$^{24}$Mg over $\sim$2~dex in metallicity,
in agreement with the distribution observed in the solar neighbourhood
\cite{F03}.

\subsection{Phosphorus}

Our new stellar models produce $^{31}$P efficiently; when coupled with the 
phosphorus associated with Type~II SNe \cite{WW95,LC03}, an
$\sim$0.2~dex enhancement in the predicted solar neighbourhood
[P/Fe] is seen across $\sim$2~dex in 
metallicity ($-$2$\simlt$[Fe/H]$\simlt$$+$0).
While empirical stellar phosphorus abundances are difficult to determine,
a firm upper limit of [P/S]$<$$+$0 has been placed on the intergalactic
medium\footnote{Or at least one damped Lyman-$\alpha$ system.} 
$\sim$2~Gyr after the Big Bang \cite{F04b}.  Such an empirical limit
was already only 0.3--0.5~dex outside the predictions of 
chemical evolution models generated without an AGB phosphorus component
(\cite{F04b}; Fig~6); the inclusion of our new $^{31}$P AGB yields
into the same damped Lyman-$\alpha$
model means the current empirical limit is less than
a factor of two outside the model predictions.  Future, more sensitive,
observations of such high-$z$ clouds can therefore, in principle, 
support or refute this specific prediction of our new LIMS models.


\begin{thebibliography}{9}
\bibitem{RV81}  A. Renzini and M. Voli, A\&A 94 (1981) 175.
\bibitem{vG97}  L.B. van~den~Hoek and M.A.T. Groenewegen, A\&AS 123 (1997) 305.
\bibitem{TWW95} F.X. Timmes, S.E. Woosley and T.A.Weaver, ApJS 98 (1995) 617.
\bibitem{KL03a} A.I. Karakas and J.C. Lattanzio, PASP 20 (2003) 279.
\bibitem{KL03b} A.I. Karakas and J.C. Lattanzio, PASP 20 (2003) 393.
\bibitem{FG03}  Y. Fenner and B.K. Gibson, PASA 20 (2003) 189.
\bibitem{G03}   B.K. Gibson, Y. Fenner, A. Renda, D. Kawata and H.-c. Lee, 
		PASA 20 (2003) 401.
\bibitem{C97}   Chiappini, C., Matteucci, F. \& Gratton, R., ApJ 477 (1997) 765
\bibitem{C01}   Chiappini, C., Matteucci, F. \& Romano, D., ApJ 554 (2001) 1044
\bibitem{F03}   Y. Fenner, B.K. Gibson, H.-c. Lee, A.I. Karakas, et~al.,
		PASA 20 (2003) 340.
\bibitem{F04a}  Y. Fenner, S. Campbell, A.I. Karakas, J.C., et al.
		MNRAS 353 (2004) 789.
\bibitem{F04b}  Y. Fenner, J.X. Prochaska and B.K. Gibson, ApJ 606 (2004) 116.
\bibitem{R04}   A. Renda, Y. Fenner, B.K. Gibson, A.I. Karakas,
		et~al., MNRAS 355 (2004) 575.
\bibitem{RM03}  D. Romano and F. Matteucci, MNRAS 342 (2003) 185.
\bibitem{WR94}  T.L. Wilson and R.T. Rood, ARAA 32 (1994) 191.
\bibitem{MM00}  G. Meynet and M. Arnould, A\&A 355 (2000) 176.
\bibitem{HL03}  J.C. Higdon and R.E. Lingenfelter, ApJ 590 (2003) 822.
\bibitem{Edv93} Edvardsson, B., Andersen, J., Gustafsson, B., et al. A\&A 275 (1993) 101
\bibitem{Gra03} Gratton, R. G., Carretta, E., Desidera, S., Lucatello, S., et al. A\&A 406 (2003) 131
\bibitem{WW95}  S.E. Woosley and T.A. Weaver, ApJS 101 (1995) 181.
\bibitem{LC03}  M. Limongi and A. Chieffi, ApJ 592 (2003) 404.
\end{thebibliography}
\end{document}